\documentclass[12pt,a4]{article}

\usepackage{theorem,amssymb,amsbsy,latexsym}

\theoremstyle{plain}

\newcommand{\mbf}[1]{{\boldsymbol {#1} }}
\newcommand{\bb}{b}
\newcommand{\opS}{{\mathbb S}}

\newcommand{\smalleq}[1]{\mbox{${#1}$}}

\newcommand{\cC}{{\cal C}}

\newcommand{\tet}{\theta}

\newcommand{\ee}[1]{{\rm e}^{#1}}
\newcommand{\ii}{{\rm i}}
\newcommand{\dd}{{\rm d}}
\newcommand{\f}{\frac}

\newcommand{\hF}{\hat{F}}

\newcommand{\vx}{{\bf x}}

\newcommand{\vn}{{\bf n}}
\newcommand{\vm}{{\bf m}}

\newcommand{\vE}{{\bf E}}
\newcommand{\vmu}{\underline{{\mbf \mu}}}

\newcommand{\vzero}{\underline{{\mbf 0}}}

\newcommand{\Ref}[1]{(\ref{#1})}
\newcommand{\binom}[2]{\left(\!\!\bma{c} {#1}\\ {#2}\ema\!\!\right)}

\newcommand{\eps}{\varepsilon}
\newcommand{\half}{\mbox{$\frac{1}{2}$}}

\newcommand{\Z}{{\mathbb Z}}

\newcommand{\cO}{{\cal O}}

\newcommand{\cE}{{\cal E}}

\newcommand{\eq}{\begin{equation}}
\newcommand{\eqend}{\end{equation}}
\newcommand{\eqa}{\begin{eqnarray}}
\newcommand{\nonueqa}{\begin{eqnarray*}}
\newcommand{\eqaend}{\end{eqnarray}}
\newcommand{\nonueqaend}{\end{eqnarray*}}
\newcommand{\nonu}{\nonumber \nopagebreak \\ \nopagebreak}
\newcommand{\bma}[1]{\begin{array}{#1}}
\newcommand{\ema}{\end{array}}
\newcommand{\bc}{\begin{center}}
\newcommand{\ec}{\end{center}}

\newcounter{saveeqn}
\newcounter{App} %\setcounter{App}{0}

\newcommand{\alpheqn}{%
\stepcounter{equation}%this is what I added
\setcounter{saveeqn}{\value{equation}}%
\setcounter{equation}{0}%
\renewcommand{\theequation}{\arabic{saveeqn}\alph{equation}} }
\newcommand{\reseteqn}{\setcounter{equation}{\value{saveeqn}}%
\renewcommand{\theequation}{\arabic{equation}} }
\newcounter{asaveeqn}

\begin{document}
\begin{flushright}
\today
\end{flushright}
\vspace{.4cm}

\begin{center}

{\Large \bf An explicit solution of the (quantum) elliptic
Calogero-Sutherland model\footnote{Contribution to the conference SPT
2004 in Cala Gonone (Sardinia, Italy)}}
\vspace{1 cm}

{\large Edwin Langmann}\\
\vspace{0.3 cm} {\em Mathematical Physics, Department of Physics, KTH,
AlbaNova, SE-106 91 Stockholm, Sweden}

\end{center}

\begin{center} 
{\em Dedicated to the memory of Ludwig Pittner} 
\end{center}

\begin{abstract}
We present explicit formulas for the eigenvalues and eigenfunctions of
the elliptic Calogero-Sutherland (eCS) model as formal power series to
all orders in the nome of the elliptic functions, for arbitrary values
of the (positive) coupling constant and particle number. Our solution
gives explicit formulas for an elliptic deformation of the Jack
polynomials.
\bigskip

\end{abstract}

\section{Introduction} The elliptic Calogero-Sutherland
(eCS) system is a quantum mechanical model of identical particles
moving on a circle and interacting via a two-body potential given by the
Weierstrass elliptic function $\wp$ \cite{C,Su,OP}. It is defined by
the 2nd order differential operator
\eq
\label{eCS}
H = - \sum_{j=1}^N\frac{\partial^2}{\partial x_j^2} \; + \; \gamma
\!\! \sum_{1\leq j<k\leq N} V(x_j-x_k) \eqend
where $N=2,3,\ldots$ is the particle number, $-\pi\leq x_j\leq \pi$
are coordinates on the circle,
\eq
\gamma = 2 \lambda (\lambda-1),\quad \lambda>0, 
\eqend
is the coupling constant, and the two-body potential
\eq
V(r) = \sum_{m\in\Z}\f{1}{ 4\sin^2[(r+ \ii\beta m )/2]} \: , \quad
\beta>0 ,
\eqend
which is essentially equal to the Weierstrass elliptic function $\wp$
with periods $2\pi$ and $\ii \beta$.\footnote{To be precise:
$V(z)=\wp(z)+c_0$ with $c_0 = 1/12 -(1/2)\sum_{m=1}^\infty
\sinh^{-2}[(\beta m)/2]$.}

The eCS system is known to be integrable in the sense that there exist
differential operators of the form
$$
H_n = \sum_{j=1}^N (-\ii)^n \frac{\partial^n}{\partial x_j^n} + \mbox{ lower order terms}
$$
for all $n=1,2,\ldots,N$, which include the eCS Hamiltonian, $H_2=H$,
and which all mutually commute, $[H_n,H_m]=0$ for $n,m=1,2,\ldots, N$
\cite{OP}.  Moreover, in the trigonometric limit $\beta=\infty$ where
the two-body potential reduces to a trigonometric function, the
explicit solution of this model was found by Sutherland more than 30
year ago \cite{Su}. In the two-particle case, $N=2$, the eigenvalue
equation of the eCS system is equivalent to the Lam\'{e} equation
studied extensively at the end of the 19th century; see \cite{WW} for
a review of the classical results. Recent work on the eCS model include  
\cite{DI,EK,EFK,FV1,FV2,P,S,T}.

In this paper we present a generalization of Sutherland's solution to
the elliptic case without restrictions on parameters (see the Result
in the final section).  More specifically, we present explicit
formulas for the eigenfunctions $\psi(\vx;\vn)$ and corresponding
eigenvalues $\cE(\vn)$ of the eCS Hamiltonian,
\eq H\psi(\vx;\vn) = \cE(\vn) \psi(\vx;\vn), \quad
\vx=(x_1,\ldots,x_N), \label{Hpsi=Epsi} \eqend
which are labeled by integer quantum numbers
\eq
\vn = (n_1,\ldots,n_N),\quad n_1\geq n_2\geq \ldots \geq n_N , \label{vn}
\eqend
and which are of the following form,
\eq \psi(\vx;\vn) = \Phi(\vx;\vn) \Psi(\vx)  \label{psi} \eqend
where 
\eq
\Psi(\vx) = \prod_{1\leq j<k\leq N}
\tet(x_j-x_k)^\lambda \label{Psi} 
\eqend
with 
\eq 
\label{tet}
\tet(z) =  \sin(z/2) \prod_{n=1}^\infty
(1-2q^{2n}\cos(z) + q^{4n})\: , \quad q=\ee{-\beta/2} 
\eqend 
essentially the Jacobi Theta function $\vartheta_1$.\footnote{To be
precise: $\tet(z) = \vartheta_1(z/2)/[2 q^{1/4} \prod_{n=1}^\infty
(1-q^{2n})]$.} The $\Phi$ are symmetric functions of the variables
$z_j=\ee{\ii x_j}$, and they are in one-to-one correspondence with the
plane waves
\eq S(\vx;\vn) = \sum_{Q\in S_N} \prod_{j=1}^N \ee{\ii n_{Qj} x_{j} }
\eqend
which provide a complete set of eigenfunctions in the non-interacting
case $\gamma=0$. It is important to note that our solution in the
trigonometric limit reduces to Sutherland's: the eigenvalues
$\cE(\vn)$ become equal to the well-known expressions
\eq \cE_0(\vn) = \sum_{j=1}^N \Bigl( n_j +
\lambda[\half(N+1)-j]\Bigr)^2 , \label{cE0} \eqend
and the functions $\Phi(\vx;\vn)$ reduce to the Jack polynomials
playing a prominent role also in various other contexts in
mathematics; see \cite{McD,St}. We also note that, for $\beta<\infty$,
the functions $\Phi(\vx;\vn)$ are no longer polynomials, and also the
eigenvalues $\cE(\vn)$ become much more complicated. In particular,
$\Psi$ is {\em not} the ground state of the eCS Hamiltonian for
$\beta<\infty$. Correspondingly, our solution is by infinite series which,
at this point, are only formal: we leave open the important but
difficult questions of convergence and resonances (as discussed in
more detail below). We only mention the results in Ref.\ \cite{KT}
which suggest that our series solutions have a finite radius of
convergence in the nome $q$ of the elliptic functions, and our results
suggest that there exists a resummation so that the resonances
disappear \cite{EL4}. Moreover, in the trigonometric limit resonances
do not appear \cite{EL2}, and all our infinite series collapse to
finite ones.

The present paper is based on our previous results in
\cite{CL,EL1,EL2,EL3,EL4}. Our starting point is a Theorem obtained by
quantum field theory techniques in \cite{EL2} and proven by direct
computations in \cite{EL4}. This theorem suggests to write the
eigenfunctions $\Phi(\vx;\vn)$ as series of particular symmetric
functions $\hF(\vx;\vn)$ which are given by the following explicit
formulas,
\eq \hF(\vx;\vn) = \prod_{j=1}^N\Bigl[
\oint_{\cC_j} \frac{\dd \xi_j}{2\pi\xi_j} \xi_j^{n_j}  \Bigr]
\frac{\prod_{1\leq j<k\leq N}\Theta(\xi_j/\xi_k)^\lambda}
{\prod_{j,k=1}^N\Theta(\ee{\ii x_j}/\xi_k)^\lambda} \eqend
where
\eq
\Theta(\xi) = (1-\xi)\prod_{m=1}^\infty [(1-q^{2m}\xi)(1-q^{2m}/\xi)], 
\eqend
and the integration contours are nested circles in the complex plane
enclosing the unit circle,
\eq \cC_j: \xi=\ee{\eps j} \ee{\ii y_j},\quad -\pi\leq y_j\leq \pi,\quad
0<\eps<\beta/N . \eqend
Note that the $\hF(\vx;\vn)$ are symmetric functions of the variables
$z_j=\ee{\ii x_j}$, and they can be expanded as Laurent series.  A
simple but important consequence of the above mentioned theorem is the
following.

\bigskip

\noindent {\bf Corollary:} {\em Let
\eq
\Phi(\vx;\vn) = \sum_{\vm} \alpha(\vmu;\vn) \hF(\vx;\vn+\vmu) 
\label{Phi} 
\eqend
where the sum is over all 
\eq \vmu = \sum_{1\leq j<k\leq N} \mu_{jk} \vE_{jk} \; \mbox{ with }
\; \mu_{jk}\in\Z \; \mbox{ and } \; (\vE_{jk})_\ell =
\delta_{j\ell}-\delta_{k\ell} \label{vmu} \eqend
for $\ell=1,2,\ldots, N$, and
\eq \alpha(\vmu;\vzero) = \delta(\vmu,\vzero) + \cO(\gamma) .  \eqend
Then $\psi(\vx;\vn)$ defined in Eqs.\ \Ref{psi} and \Ref{Psi} is an
eigenfunction of the eCS Hamiltonian with corresponding eigenvalue
$\cE(\vn)$ provided that
\eq [\cE_0(\vn+\vmu) -\cE(\vn) ] \alpha(\vmu;\vn) = \gamma\sum_{j<k}
\sum_{\nu\in\Z} S_\nu \alpha (\vmu - \nu \vE_{jk};\vn) \label{Eq} \eqend
with $\cE_0(\vn)$ in Eq.\ \Ref{cE0} and
\eq S_0=0 , \quad S_\nu = \frac{\nu}{1-q^{2\nu}}\quad \mbox{ and }
\quad S_{-\nu} = \frac{\nu q^{2\nu}}{1-q^{2\nu}} \quad \mbox{ for
$\nu>0$} .  \eqend
} 

\bigskip

\noindent Note that $S_\nu = [1-\delta(\nu,0)]\nu/(1-q^{2\nu})$ for
all integer $\nu$, but we prefer our somewhat less elegant definition
which makes manifest that, in the trigonometric limit $q=0$, $S_\nu=0$
for $\nu\leq 0$. This implies that, for $q=0$, Eq.\ \Ref{Eq} has
triangular structure (in a natural sense explained in Ref.\
\cite{EL2}), which implies that the eigenvalue $\cE(\vn)$ is equal to
$\cE_0(\vn)$. Moreover, one obtains a simple recursion relation to
compute all the non-zero $\alpha(\vmu;\vn)$ in in Eq.\ \Ref{Phi}. It
is interesting to note that this solution algorithm for $q=0$ is
different from Sutherland's, even though is yields the same solution
\cite{EL2}. We did not realize in \cite{EL2} that it is possible to
obtain an explicit solution of the recursion relations for the
coefficients $\alpha(\vmu;\vn)$ as follows,
\eqa \alpha(\vmu;\vn) = \delta(\vmu;\vzero) +
\sum_{s=0}^{\infty}\gamma^s \sum_{j_1<k_1}\cdots
\sum_{j_s<k_s}\sum_{\nu_1,\ldots,\nu_s=1}^\infty \nonu \times
\frac{\nu_1\nu_2\cdots\nu_s \delta(\vmu,\sum_{r=1}^s \nu_r
\vE_{j_r,k_r})}{ \prod_{r=1}^s \bb(\sum_{\ell=1}^r
\nu_\ell\vE_{j_\ell,k_\ell};\vn)} \quad \mbox{ for $q=0$}, \label{q=0}
\eqaend
with 
\eq \bb(\vmu;\vn) = \cE_0(\vn+\vmu) -\cE_0(\vn) \label{bb} , \eqend
where all sums are in fact finite due to the Kronecker delta and
certain properties of the functions $\hF(\vx;\vn)$ discussed in
\cite{EL2} (Eq.\ \Ref{q=0} is a simple special case of our general
result presented in the last section).  This together with the results
in Ref.\ \cite{EL2} provides explicit formulas for the Jack
polynomials.  It is interesting to note that the recursions relations
which one gets in Sutherland's algorithm are more complicated and, to
our knowledge, have not been solved explicitly for general $N$ (we are
only aware of similarly explicit previous results for $N=3$
\cite{PRZ}). It is also worth noting that, due to translation
invariance, the dependence of the eigenfunctions on the center-of-mass
coordinate $x_1+\cdots + x_N$ is trivial, and therefore the functions
$\Phi(\vx;\vn)$ are of the form $\ee{i n_N(x_1+ \cdots x_N)}
\Phi(\vx;\tilde\vn)$ where $\tilde n_j=n_j-n_N$ obeys $\tilde n_1\geq
\tilde n_2\geq \ldots \tilde n_{N-1} \geq \tilde n_N=0$. However, it
seems that this is not manifest in our explicit formula for these
functions, and we therefore seem to get an infinite number of
different representations (labeled by $n_N$) for each distinct Jack
polynomial (this is true not only for $q=0$ but also in the elliptic
case).

In the rest of the paper we discuss how to explicitly compute the
coefficients $\alpha(\vmu;\vn)$ and eigenvalues $\cE(\vn)$ for the
general elliptic case from the Corollary above.  The strategy is to
find a ``good'' expansion parameter, allowing to solve Eq.\ \Ref{Eq}
recursively. The obvious parameter is the squared nome $q^2$ of the
elliptic function \cite{EL1}. While this provides a possible solution
algorithm, we were able to obtain the explicit solution only up to
order $(q^{2})^7$ for $N=2$ with the help of MAPLE in this way
\cite{EL4}. The formulas obtained are rather complicated and suggest
that it is hopeless to find explicit expressions to all order in
$q^2$. However, this result motivates a more efficient solution
strategy by expanding in the coupling parameter $\gamma$. As we show,
the resulting solution algorithm is indeed simpler, and while we
obtained the explicit solution up to order $\gamma^6$ {\em without}
the help of MAPLE, it again seems hopeless to obtain explicit formulas
to all orders in $\gamma$ in that way. However, this result gives a
better understanding of the structure of the solutions, and it led us
to a method of solution to all order. The key to this was to introduce
a parameter $\eta$ ``by hand'' which efficiently organizes the
complexity of the solution, and this allows us to obtain the solution
as a power series in $\eta$ to all orders (without the help of
MAPLE). From this our explicit formulas for $\cE(\vn)$ and
$\alpha(\vmu;\vn)$ of Eq.\ \Ref{Eq} are obtained by setting $\eta=1$.

In the next three sections we outline the explicit solutions obtained
by expanding in $q^2$, $\gamma$ and $\eta$, respectively. For
simplicity in notation we restrict this discussion to the simplest
non-trivial case $N=2$. The generalization to arbitrary $N$ is
straightforward, but we only give the result in the final section.  We
plan to include a more detailed derivation of this in a future
revision of Ref.\ \cite{EL4}.

\section{Expanding in $q^2$} 
For simplicity we restrict ourselves to $N=2$. In this case
$\vmu=\mu\vE_{12}$, and \Ref{Eq} simplifies to
\eq [\cE_0(\vn + \mu\vE_{12})  - \cE(\vn)] \alpha(\mu;\vn) = \gamma
\sum_{\nu\in\Z} S_\nu \alpha (\mu - \nu;\vn) \label{Eq2} \eqend
where $\alpha(\mu;\vn)$ is short for $\alpha(\mu\vE_{12};\vn)$. To
simplify notation we suppress the dependence on $\vn$ in the
following.  We make the ansatz
\eq \alpha(\mu) = \sum_{\ell=0}^\infty \alpha_\ell(\mu)
q^{2\ell},\quad \cE = \sum_{\ell=0}^\infty \cE_\ell
q^{2\ell} , \eqend
and with
\eq \alpha_\ell(0) = \delta(\mu,0)\quad \mbox{ and } \quad
\alpha_\ell(\mu) = 0 \mbox{ for } \mu<-\ell  \label{hw} \eqend
we obtain by simple computations (expanding the $S_\nu$ in geometric
series etc.)
\eqa \bb(\mu) \alpha_\ell(\mu) =
\sum_{m=1}^\ell\cE_m\alpha_{\ell-m}(\mu) + \gamma\sum_{\nu=1}^\ell \nu
\alpha_\ell(\mu-\nu) \nonu + \gamma \sum_{\nu=1}^\ell
\sum_{m=1}^{\ell/\nu} \nu[ \alpha_{\ell-\nu m}(\mu-\nu) +
\alpha_{\ell-\nu m}(\mu + \nu) ] \label{al1} \eqaend
where $\bb(\mu) = \cE_0(\vn+\mu\vE_{12})-\cE_0(\vn)$. Using 
$\cE_0(\vn)=(n_1+\lambda/2)^2+(n_2-\lambda/2)^2$ we get
\newcommand{\tn}{P} 
\eq
\bb(\mu) = 2\mu(\tn+\mu) ,\quad \tn = n_1-n_2 +\lambda .  
\eqend
It is important to note that Eq.\ \Ref{al1} has triangular structure:
for each $\ell$ and $\mu\neq 0$ it determines $\alpha_\ell(\mu)$ as a
finite sum of terms involving only $\alpha_{\ell'}(\mu')$ and
$\cE_{\ell'}$ with $\ell'<\ell$ and $\alpha_\ell(\mu')$ with
$\mu'<\mu$, and the equation for $\mu=0$ allows to determine the
$\cE_\ell$ recursively.  We also note that the conditions in Eq.\
\Ref{hw} is essential for getting a simple recursion procedure. It is
straightforward to implement the recursive computation of the
$\alpha_\ell(\mu)$ and $\cE_\ell$ in a symbolic computing software
like MAPLE or MATHEMATICA. We only give the results for the
eigenvalues which we obtained using MAPLE,
\alpheqn 
\eqa \cE_1 &=& \frac{1}{\tn^2-1}\gamma^2 \label{E1} \\
\cE_2 &=&
\frac{1}{(\tn^2-4)(\tn^2-1)}\Bigl[ \smalleq{6(\tn^2-2)\gamma^2 - 6\gamma^3 +
\frac{(5\tn^2+7)}{4(\tn^2-1)^2}\gamma^4 } \Bigr] \label{E2} \\
\cE_3 &=& \frac{1}{(\tn^2-9)(\tn^2-1)}\Bigl[ \smalleq{ 12
(\tn^2-3)\gamma^2 -48\gamma^3 +\frac{4(15\tn^4 -37\tn^2
-2)}{(\tn^2-4)(\tn^2-1)^2}\gamma^4 } \nonu && 
\smalleq{\qquad\qquad\qquad\qquad  - 
\frac{4(7\tn^2 +17)}{(\tn^2-4)(\tn^2-1)^2}\gamma^5 + \frac{(9\tn^4
+58\tn^2 +29) }{2(\tn^2-4)(\tn^2-1)^4}\gamma^6 } \Bigr] \label{E3} \\ 
\cE_4 &=& \frac{1}{(\tn^2-16)(\tn^2-1)}\Bigl[ \smalleq{
4\frac{(7\tn^4-74\tn^2+112)}{(\tn^2-4)}\gamma^2 -180\gamma^3 } \nonu &&
\smalleq{\qquad\qquad\qquad\qquad  +
\frac{3(365\tn^{10}-6662\tn^8+42249\tn^6-115640\tn^4+119816\tn^2-18528)}{2(\tn^2-9)(\tn^2-2)^2(\tn^2-1)^2}\gamma^4
}\nonu && \smalleq{ \qquad\qquad\qquad\qquad -
\frac{3(259\tn^8-3358\tn^6+11415\tn^4-4252\tn^2-25664)}{(\tn^2-9)(\tn^2-4)^3(\tn^2-1)^2}\gamma^5
} \\ && \smalleq{\qquad\qquad\qquad\qquad  +
\frac{2151\tn^{10}-18127\tn^8-10529\tn^6+293115\tn^4-501962\tn^2+79832}{4(\tn^2-9)(\tn^2-4)^3(\tn^2-1)^4}\gamma^6
}\nonu && \smalleq{\qquad\qquad\qquad\qquad\qquad  -
\frac{715\tn^8-481\tn^6-43203\tn^4+94061\tn^2+104428}{4(\tn^2-9)(\tn^2-4)^3(\tn^2-1)^4}\gamma^7
} \nonu && \smalleq{\qquad\qquad\qquad\qquad  +
\frac{1469\tn^{10}+9144\tn^8-140354\tn^6+64228\tn^4+827565\tn^2+274748}{64(\tn^2-9)(\tn^2-4)^3(\tn^2-1)^6}\gamma^8}
\Bigr] \: .  \label{E4} \nonumber 
\eqaend
\reseteqn We computed $\cE_\ell$ up to order $\ell=7$ but for obvious
reasons do not write down our full result. This result was previously
obtained in \cite{P} up to $\ell=2$ using a different method, and we
convinced ourselves that our results agree.

\bigskip

\noindent {\bf Remark:} In the computations discussed above we
implicitly assumed that $\bb(\mu)$ is always different from zero,
which is only the case if $\lambda$, and thus $\tn$, is not an
integer. If $\bb(\mu)$ vanishes we have a {\em resonance}, and while
it is possible to generalize the algorithm to allow for resonances
\cite{EL4} we will not discuss this here for simplicity: Throughout
this paper we ignore resonances. (As discussed in \cite{EL4}, we
believe that resonances are not a serious problem.)                         .)

\bigskip

We observe that the $\cE_\ell$ become more and more complicated with
increasing $\ell$, and it seems that they are polynomials in the
coupling constant $\gamma$ as follows,
\eq \cE_\ell =
\sum_{s=2}^{2\ell}\cE^{(s)}_{\ell} \gamma^s \eqend
where $\cE^{(s)}_\ell$ become more complicated with increasing $s$.
It is interesting to note that the formulas for the coefficients
$\cE^{(s)}_\ell$ become somewhat simpler when expanded in partial
fractions. In particular, we found by inspection that all
$\cE^{(2)}_\ell$ we computed can be written in the following simple
form,
\eq \cE^{(2)}_{\ell} = \frac{\ell}2 \sum_{k \vert \ell} \left(
\frac1{(P-k)} -\frac1{(P+k)} \right) \label{conj1} \eqend
(the sum is over all integer divisors $k$ of $\ell$). We checked this
formula up to $\ell=7$, and by assuming it to be true for all $\ell$ we
obtain by a simple computation
\eq \cE^{(2)} = \sum_{\ell=0}^\infty \cE^{(2)}_{\ell} q^{2\ell} =
\sum_{k=1}^\infty \left( \frac1{P-k} - \frac1{P+k} \right) \frac{k
q^{2k}}{2 (1-q^{2k})^2 } \label{cE2c} \eqend
(this conjecture will be proven in the next section). This formula
suggests that it is possible to obtain explicit expressions to all
orders in $q^2$ if one expands in the coupling parameter $\gamma$.  We
now present an alternative and simpler solution algorithm motivated by
this observation.

\section{Expanding in $\gamma$}
We now make the ansatz
\eq \alpha(\mu) = \sum_{s=0}^\infty \alpha^{(s)}\gamma^s,\quad \cE =
\cE_0 + \sum_{s=1}^\infty \cE^{(s)}\gamma^s, \eqend
and by simple computations we obtain from Eq.\ \Ref{Eq} 
\eq \bb(\mu) \alpha^{(s)}(\mu) - \sum_{r=0}^s\cE^{(s-r)}
\alpha^{(r)}(\mu) =  \sum_{\nu\in\Z} S_\nu \alpha^{(s-1)}(\mu-\nu) \label{al2} \eqend
which we can solve with the following ansatz
\eq \alpha^{(0)}(\mu) = \delta(\mu,0),\quad \alpha^{(s)}(0)=0 \mbox{
for } s=1,2,\ldots .  \eqend
This yields, in particular,
\eq
\cE^{(s)} = -\sum_{\nu\in\Z} S_\nu \alpha^{(s-1)}(-\nu) . \label{Es} 
\eqend
The recursion relations in Eqs.\ \Ref{al2} are much simpler than the
ones in Eq.\ \Ref{al1}. By inspection we find that the following
ansatz is consistent,
\eq
\alpha^{(s)}(\mu) = \sum_{\nu_1,\ldots,\nu_s\in\Z} S_{\nu_s} \cdots  S_{\nu_1} 
f_s(\nu_1,\ldots,\nu_s;\mu) , 
\eqend
and by straightforward computations,
\alpheqn \eqa f_1(\nu_1;\mu) &=&
\frac{\delta(\mu,\nu_1)}{\bb(\nu_1)}\\ f_2(\nu_1,\nu_2;\mu) &=&
\frac{\delta(\mu,\nu_1+\nu_2)}{\bb(\nu_1+\nu_2)\bb(\nu_1) }\\
f_3(\nu_1,\nu_2,\nu_3;\mu) &=&
\frac{\delta(\mu,\nu_1+\nu_2+\nu_3)}{\bb(\nu_1+\nu_2+\nu_3)
\bb(\nu_1+\nu_2)\bb(\nu_1) } -
\frac{\delta(\nu_2+\nu_3,0)\delta(\mu,\nu_1)}{\bb(\nu_2)\bb(\nu_1)^2}
\\ f_4(\nu_1,\nu_2,\nu_3,\nu_4;\mu) &=&
\frac{\delta(\mu,\nu_1+\nu_2+\nu_3+\nu_4)}{\bb(\nu_1+\nu_2+\nu_3+\nu_4)\bb(\nu_1+\nu_2+\nu_3)
\bb(\nu_1+\nu_2)\bb(\nu_1) } \nonu && -
\frac{\delta(\nu_2+\nu_3,0)\delta(\mu,\nu_1+\nu_4)}{\bb(\nu_2)\bb(\nu_1)^2\bb(\nu_1+\nu_4)}
-
\frac{\delta(\nu_3+\nu_4,0)\delta(\mu,\nu_1+\nu_2)}{\bb(\nu_4)\bb(\nu_1+\nu_2)^2\bb(\nu_1)}
\nonu && -
\frac{\delta(\nu_2+\nu_3+\nu_4,0)\delta(\mu,\nu_1)}{\bb(\nu_3+\nu_4)\bb(\nu_3)\bb(\nu_1)^2}
\eqaend \reseteqn
etc.\ (it is useful to note that the variables $\nu_j$ can be permuted
in each term). Using Eq.\ \Ref{Es} we get
\eq \cE^{(s)} =-\sum_{\nu_1,\ldots,\nu_s\in\Z} S_{\nu_s}\cdots
S_{\nu_1} f_{s-1}(\nu_{1}, \ldots, \nu_{s-1};-\nu_s),  \eqend
in particular, $\cE^{(1)}=0$ and
\eq \cE^{(2)} =
-\sum_{\nu_2,\nu_1}S_{\nu_2}S_{\nu_1}\frac{\delta(\nu_1+\nu_2,0)}{\bb(\nu_1)}
= -\sum_{\nu=1}^\infty S_\nu S_{-\nu} \Bigl( \frac1{\bb(\nu)} +
\frac1{\bb(- \nu)}\Bigr) \eqend
which proves Eq.\ \Ref{cE2c}. With the formulas given above one can
easily write down similarly explicit formulas for $\cE^{(s)}$ for
$s=2,3,4,5$.

Computing the functions $f_s$ up to $s=6$ we observed the following
simple patterns: it seems that the building blocks for the solution
are the following expressions,
\eq G^{(s)}_\ell = \sum_{\nu_1,\ldots,\nu_s\in\Z} S_{\nu_s} \cdots 
S_{\nu_1} \sum_{k_1,\ldots,k_{s-1}=0}^\infty 
\delta(k_1+\cdots +k_{s-1} -\ell) \frac{\delta(\sum_{\ell=1}^s
\nu_s,0)}{\prod_{r=1}^{s-1}\bb(\sum_{\ell=1}^r \nu_\ell)^{1+k_r}} . 
\eqend
With that we can write the coefficients of the eigenvalues in a simple
manner as follows,
\alpheqn
\eqa
\cE^{(2)} &=& -G^{(2)}_0\\
\cE^{(3)} &=& -G^{(3)}_0\\
\cE^{(4)} &=& -G^{(4)}_0 +G^{(2)}_0 G^{(2)}_1  \\
\cE^{(5)} &=& -G^{(5)}_0 +G^{(3)}_0 G^{(2)}_1  + G^{(2)}_0 G^{(3)}_1 \\ 
\cE^{(6)} &=& -G^{(6)}_0 +G^{(4)}_0 G^{(2)}_1  + G^{(2)}_0 G^{(4)}_1 \nonu 
&& + G^{(3)}_0 G^{(3)}_1 - G_0^{(2)} [G_1^{(2)}]^2 - [G_0^{(2)}]^2 G_2^{(2)}  . \label{Ess} 
\eqaend
\reseteqn
This suggests that 
\eq \cE^{(s)} = \sum_{r=1}^\infty (-1)^r (\cdots) \delta(s_1+\cdots
s_r,s) \delta(k_1+\cdots + k_r,r-1) G^{(s_1)}_{k_1} \cdots
G^{(s_r)}_{k_r} \label{cj}
\eqend
where `$(\cdots)$' are possible combinatorial factors which might
appear at higher order but, up to $s=6$, all are equal to 1.

We find that, for $s>6$, nontrivial combinatorial factors in the
formula above appear, and this destroys the hope that we can find a
closed formula for $\cE^{(s)}$ for arbitrary $s$ in this way.  Still,
Eq.\ \Ref{cj} suggest that we can write $\cE$ in a simple manner using
the following quantities,
\eq
G_k = \sum_{s=2}^\infty G_k^{(s)} \gamma^s , 
\eqend
and the formulas above suggest,
\eq \cE = \cE_0 - G_0 + G_0 G_1 - G_0[G_1]^2 - [G_0]^2G_2 + \ldots \label{conjecture} 
\eqend
where the dots are higher order terms. We will prove and extend this
formula in the next section: We will obtain an explicit formula of the
following kind,
\eq \cE = \cE_0 + \sum_{n=1}\tilde \cE_n,\quad \tilde \cE_n =
\sum_{r_1,\ldots,r_n} (\cdots) G_{r_1}\cdots G_{r_n} \eqend
with certain combinatorial factor `$(\cdots)$' which we will compute
explicitly.  It is interesting to note that $\tilde \cE_n$ is of order
$q^{2n}$ (since all $G_s$ are of order $q^2$), and thus this formula
allows to deduce in a simple manner the series expansion in
$q^2$. However, the term $\tilde \cE_n$ contributes to all the
$(q^2)^m$-terms $m\geq n$, which explains why our expansion in $q^2$
yielded so complicated expressions. We stress that, since
$\cE_n=\cO(q^{2n})$, Eq.\ \Ref{cE1} still is an expansion the $q^2$
and has, as we believe, a finite radius of convergence.

\section{Expanding in $\eta$} 
The results in the previous section led us to a more powerful solution
strategy which we now explain. Defining an operator $\opS$ as follows,
\eq
\opS f(\mu) \, :=\,  \sum_{\nu\in\Z} S_\nu f(\mu-\nu)
\eqend
we can write Eq.\ \Ref{Eq} as
\eq
[\bb(\mu) - \tilde\cE] \alpha(\mu) = \gamma \opS \alpha(\mu) \label{Eq3} 
\eqend
where
\eq \tilde\cE = \cE-\cE_0 .  \eqend
Making the ansatz 
\eq
\alpha(\mu) = \delta(\mu,0) + \sum_{s=1}^\infty \tilde \alpha_s(\mu)\gamma^s
\eqend
we get $[\bb(\mu)-\tilde\cE]\tilde \alpha_s(\mu)=\opS\tilde
\alpha_{s-1}(\mu)$, and thus
\eqa \alpha(\mu) = \sum_{s=0}^\infty \gamma^s\Bigl( [\bb(\mu)-\tilde\cE]^{-1}
\opS\Bigr)^s\delta(\mu,0) = \nonu \sum_{s=0}^\infty \gamma^s 
\sum_{\nu_1,\ldots,\nu_s\in\Z}
S_{\nu_s}\cdots S_{\nu_1} \frac1{[\bb(\mu)-\tilde\cE]}
\frac1{[\bb(\mu-\nu_{s} )-\tilde\cE]} \frac1{[\bb(\mu-\nu_{s} -\nu_{s-1}
)-\tilde\cE]} \nonu \cdots \frac1{[\bb(\mu-\nu_{s} -\cdots -\nu_{2}
)-\tilde\cE]} \delta(\mu-\nu_s-\nu_{s-1}-\cdots - \nu_1,0) = \nonu
\sum_{s=0}^\infty \gamma^s 
\sum_{\nu_1,\ldots,\nu_s\in\Z} S_{\nu_s}\cdots S_{\nu_1}
\frac{\delta(\mu,\sum_{r=1}^s\nu_r)} { \prod_{r=1}^{s}
[\bb(\mu-\sum_{\ell=r+1}^s\nu_\ell)-\tilde \cE]} . \eqaend
Setting $\mu=0$ in Eq.\ \Ref{Eq3} gives
$\tilde\cE=-\gamma\opS\alpha(0)$, which implies
\eq
\tilde \cE = -\sum_{s=0}^\infty \gamma^{s+1} 
\sum_{\nu_1,\ldots,\nu_{s+1} \in\Z} S_{\nu_{s+1} }\cdots S_{\nu_1}
 \frac{\delta(\sum_{r=1}^{s+1}\nu_r,0)} { \prod_{r=1}^{s}
[\bb(\mu-\sum_{\ell=r+1}^{s+1}\nu_\ell)-\tilde \cE]} .
\eqend
It is easy to see that the $s=0$ term here vanishes, and by a shifting
the summation variable we obtain the following equation determining
$\tilde\cE$,
\eq
\tilde \cE  = -G(\tilde\cE) \label{cE1} 
\eqend
where
\eq G(\xi) \,:=\, \sum_{s=2}^\infty \gamma^{s} \sum_{\nu_1,\ldots,\nu_{s}
\in\Z} S_{\nu_{s} }\cdots S_{\nu_1}
\frac{\delta(\sum_{r=1}^{s}\nu_r,0)} { \prod_{r=1}^{s-1}
[\bb(\sum_{\ell=1}^{r}\nu_\ell)-\xi]} \label{G} .  \eqend
We now observe that the functions $G(\xi)$ has a Taylor expansion as follows,
\eq
G(\xi) = \sum_{k=0}^\infty G_k\xi^k
\eqend
with
\eqa G_k = \frac1{k!} \left . \frac{\dd^k}{\dd \xi^k}
\sum_{s=2}^\infty \gamma^{s} \sum_{\nu_1,\ldots,\nu_{s} \in\Z}
S_{\nu_{s} }\cdots S_{\nu_1} \frac{\delta(\sum_{r=1}^{s}\nu_r,0)} {
\prod_{r=1}^{s-1} [\bb(\sum_{\ell=1}^{r}\nu_\ell)-\xi]}\right|_{\xi=0}
\nonu = \sum_{s=1}^\infty \gamma^{s} \sum_{\nu_1,\ldots,\nu_{s} \in\Z}
\prod_{r=1}^s S_{\nu_{r} } \sum_{k_1,\ldots,k_s=0}^\infty
\delta(k,\smalleq{\sum_{r=1}^s k_r})
\frac{\delta(\smalleq{\sum_{r=1}^{s}\nu_r,0)} } { \prod_{r=1}^{s-1}
[\bb(\sum_{\ell=1}^{r}\nu_\ell)-\xi]^{1+k_s} }.  \eqaend
Note that these are exactly the quantities which we found empirically
in the last Section.  To solve Eq.\ \Ref{cE1} efficiently we replace
it by
\eq \tilde \cE = -\eta \sum_{k=0}^\infty G_k (\tilde \cE)^k \label{c0}
\eqend
where we introduce a parameter $\eta$ serving as useful book keeping
device (we set $\eta=1$ at the end of the computation). It is
straightforward to compute the Taylor series of $\tilde \cE$
recursively: With the Ansatz
\eq
\tilde \cE=\sum_{k=1}^\infty \tilde \cE_k\eta^k\label{c1} 
\eqend
we get by simple computations 
\eqa
 \tilde\cE_1 &=& -G_0\nonu
 \tilde\cE_2 &=&  G_0G_1\nonu
 \tilde\cE_3 &=& -[G_0]^2 G_2- G_0[G_1]^2 \nonu
 \tilde\cE_4 &=& 3[G_0]^2 G_1 G_2 + G_0 [G_1]^3 + [G_0]^3 G_3\nonu
 \tilde\cE_5 &=& -4[G_0]^3 G_1 G_3 - 6 [G_0]^2[G_1]^2G_2 -2 [G_0]^3[G_2]^2 - [G_0]^3 G_4 - G_0[G_1]^4
\eqaend
etc. This suggests that 
\eq \tilde\cE_n = (-1)^n \sum_{k_0,k_1,\cdots,k_{n-1}=0}^\infty
\delta(n-1,\smalleq{\sum_{j=1}^{n-1} jk_j })\delta(n,\smalleq{\sum_{j=0}^{n-1} 
k_j }) \binom{n-1}{k_0,\ldots,k_{n-1} } \prod_{j=0}^{n-1} [G_j]^{k_j}
\label{c2} \eqend
for all $n=1,2,\ldots$ (we checked that using MAPLE up to
$n=15$). This result is, in fact, a classical theorem due to
Lagrange:\footnote{I thank S.G.\ Rajeev and G.\ Lindblad for helpful
discussions on this.} the equation determining $\tilde\cE$ is of the
form $\tilde\cE = -\eta G(\tilde\cE)$, and thus Lagrange's theorem as
stated in \cite{WW}, Paragraph~7.32, implies
\eqa \tilde \cE = \sum_{n=1}^\infty \frac{(-\eta)^n}{n!}
\left. \frac{\dd^{n-1}}{\dd y^{n-1}} G(y)^n \right|_{y=0} \eqaend
equivalent to Eq.\ \Ref{c2}. 

\section{Conclusions}  
It is possible to generalize the results of the previous two
sections to arbitrary particle numbers $N$. We intend to give the
details in a future revision of Ref.\ \cite{EL4} and quote here only
the result.
\bigskip

\noindent {\bf Result:} {\em The eigenvalues of the elliptic eCS
model are given by
\eqa \cE(\vn) = \cE_0(\vn) + \sum_{n=1}^\infty (-1)^n
\sum_{k_0,\cdots,k_{n-1}=0}^\infty \nonu\times 
\delta(n-1,\smalleq{\sum_{j=1}^{n-1} jk_j
}) \delta(n,\smalleq{\sum_{j=1}^{n-1} k_j }) \binom{n-1}{k_0,\ldots,k_{n-1} } 
\prod_{j=0}^{n-1} [G_j(\vn)]^{k_j} \eqaend
with $\cE_0(\vn)= \sum_{j=1}^N \Bigl( n_j +
\lambda[\half(N+1)-j]\Bigr)^2$, $n_1\geq n_2\geq \ldots n_N$ integers, and
\eqa G_k(\vn) = \sum_{s=2}^\infty \gamma^{s} \sum_{j_1<k_1} \cdots
\sum_{j_s<k_s} \sum_{\nu_1,\ldots,\nu_{s} \in\Z} S_{\nu_{s} } \cdots
S_{\nu_{1} } \sum_{\ell_1,\ldots,\ell_s=0}^\infty
\nonu\times
\delta(k,\smalleq{\sum_{r=1}^s \ell_r})
\frac{\delta(\smalleq{\sum_{r=1}^{s}\nu_r\vE_{j_r k_r},\vzero)} } {
\prod_{r=1}^{s-1} \bb(\sum_{\ell=1}^{r}\nu_\ell\vE_{j_\ell
k_\ell};\vn)^{1+\ell_s} } \eqaend
where $\bb(\vmu;\vn)=\cE_0(\vn+\vmu)-\cE_0(\vn)$, $S_\nu =
[1-\delta(\nu,0)]\nu/(1-q^{2\nu})$ and $\vE_{jk}$ in Eq.\ \Ref{vmu}.
The corresponding eigenfunctions are given by Eqs.\ \Ref{psi},
\Ref{Psi} and \Ref{Phi}--\Ref{vmu} with the coefficients
\eqa
\alpha(\vmu) = \delta(\vmu,\vzero) + 
\sum_{s=1}^\infty \gamma^s \sum_{j_1<k_1}\cdots \sum_{j_s<k_s}
\sum_{\nu_1,\ldots,\nu_s\in\Z} S_{\nu_s}\cdots S_{\nu_1}\nonu\times 
\frac{\delta(\vmu,\sum_{r=1}^s\nu_r\vE_{j_r k_r} )} { \prod_{r=1}^{s}
[\bb(\sum_{\ell=1}^r\nu_\ell;\vn)- \cE(\vn)+\cE_0(\vn) ]} . 
\eqaend
}
\bigskip

There is an even more explicit formula for the coefficients
$\alpha(\vmu)$ which we plan to give elsewhere.

We do not label this as theorem since we did not check the details of
our formulas here carefully enough to be sure that there are no typos
and/or (minor) mistakes.\footnote{We finished this paper under the
pressure of a deadline. Moreover, we are aware of our unfortunate
tendency to make errors when copying formulas from notes to a LaTex
file.} The purpose of this paper was to make available the explicit
solution of the eCS model which we announced in two recent meetings,
and to describe the last part of a somewhat lengthy journey leading us
to this result. We feel that we are not at the end of this journey
yet: quite some work remains to be done to understand this result.

Anyway, it seems fair to say that the eCS model is an exactly {\em
solved} model now.

\section*{Acknowledgments}
I would like to thank V.B. Kuznetsov for helpful discussions and
Martin Halln\"as for reading the manuscript.  This work was supported
by the Swedish Science Research Council~(VR) and the G\"oran
Gustafsson Foundation.

\end{document}